\documentclass[showpacs,aps,epsfig,nofootinbib,superscriptaddress]{revtex4}
\usepackage{mathrsfs}
\usepackage{graphicx}
\usepackage{amsfonts}
\usepackage{multirow}
\usepackage{epstopdf}
\usepackage{latexsym}
\usepackage{amssymb}
\usepackage{amssymb}
\usepackage{ulem}
\usepackage{color}
\usepackage{amsmath}
\usepackage[center]{subfigure}
\usepackage{marvosym}
\usepackage{threeparttable}
\usepackage{booktabs}

\begin{document}

	\newcommand{\bq}{\begin{equation}}
	\newcommand{\eq}{\end{equation}}
	\newcommand{\bqn}{\begin{eqnarray}}
		\newcommand{\eqn}{\end{eqnarray}}
	\newcommand{\nb}{\nonumber}
	\newcommand{\lb}{\label}
	\newcommand{\PRL}{Phys. Rev. Lett.}
	\newcommand{\PL}{Phys. Lett.}
	\newcommand{\PR}{Phys. Rev.}
	\newcommand{\CQG}{Class. Quantum Grav.}
	

	\title{Machine Learning for Nanohertz Gravitational Wave Detection and Parameter Estimation with Pulsar Timing Array
}
\author{MengNi Chen}
\affiliation{College of Physics, Chongqing University,
	Chongqing 401331, China}
\author{Yuanhong Zhong$^{*}$}
\affiliation{School of Microelectronics and Communication Engineering, Chongqing University, Chongqing 400044, China}
\author{Yi Feng}
\affiliation{ National Astronomical Observatories, Chinese Academy of Sciences, Beijing 100012, China}
\affiliation{University of Chinese Academy of Sciences, Beijing 100049, China}
\affiliation{CAS Key Laboratory of FAST, National Astronomical Observatories, Chinese Academy of  Sciences, Beijing 100101, China}
\author{Di Li}
\affiliation{CAS Key Laboratory of FAST, National Astronomical Observatories, Chinese Academy of  Sciences, Beijing 100101, China}
\affiliation{NAOC-UKZN Computational Astrophysics Centre, University of KwaZulu-Natal, Durban 4000, South Africa}
\author{Jin Li$^{1}$}

\email{zhongyh@cqu.edu.cn}
\email{cqujinli1983@cqu.edu.cn}
\date{\today}

	\begin{abstract}
	Studies have shown that the use of pulsar timing arrays (PTAs) is among the approaches with the highest potential to detect very low-frequency gravitational waves in the near future. Although the capture of gravitational waves (GWs) by PTAs has not been reported yet, many related theoretical studies and some meaningful detection limits have been reported. In this study, we focused on the nanohertz GWs from individual supermassive binary black holes. Given specific pulsars (PSR J1909$-$3744, PSR J1713$+$0747, PSR J0437$-$4715), the corresponding GW$-$induced timing residuals in PTAs with Gaussian white noise can be simulated. Further, we report the classification of the simulated PTA data and parameter estimation for potential GW sources using machine learning based on neural networks. As a classifier, the convolutional neural network shows high accuracy when the combined signal to noise ratio $\geq$1.33 for our simulated data. Further, we applied a recurrent neural network to estimate the chirp mass ($\mathcal{M}$) of the source and luminosity distance ($\text{D}_{p}$) of the pulsars and Bayesian neural networks (BNNs) to obtain the uncertainties of chirp mass estimation. Knowledge of the uncertainties is crucial to astrophysical observation. In our case, the mean relative error of chirp mass estimation is less than $13.6\%$. Although these results are achieved for simulated PTA data, we believe that they will be important for realizing intelligent processing in PTA data analysis.

		
	\end{abstract}
	
\maketitle
	
	\section{Introduction}
	The use of pulsar timing arrays (PTAs) to detect nanohertz gravitational waves (GWs) is being considered as a supplement to the direct detection of GWs, this approach based on PTAs is considered a major milestone in GW astrophysics following the detection of GWs in the Laser Interferometer Gravitational-Wave Observatory (LIGO) frequency band. The target of GW detection using PTAs is to identify the influence of GW on the arrival time of pulsar signals at the Earth \cite{ref33,ref34}. The continuous waves from individual inspiralling supermassive binary black holes (SMBBHs) is among the promising GW sources in the nanohertz frequency range \cite{ref1,ref9}. In addition, there has been significant improvement in the GW signal search in PTA data \cite{ref2}. Therefore, to investigate the effectiveness of neural-network-based machine learning for PTA data analysis, we focused on the GWs from individual SMBBHs.
	
	Although no statistically significant GW has been detected as yet, the upper limits of the detectable GW strain amplitude in the PTA frequency range have been constantly improved (for example, $h\geq1.7\times 10^{-14}$ at 10$^{-8}\rm{Hz}$ at present) \cite{ref28}. Meanwhile, the sensitive values of GW source (i.e., SMBBHs) parameters were derived: chirp mass $\mathcal{M}\sim10^9 M_{\odot}$, and luminosity distance $D_{L}\sim100$ Mpc \cite{ref1}. These results were derived using the matched filtering method \cite{ref3} since the signal model was clear. In this study, we adopted a completely different data analysis method, namely, machine learning based on neural networks, to identify GW signals and estimate the chirp masses of SMBBHs, and the luminosity distances of the specific pulsars.
	
	With the development of intelligent data-processing technologies, machine learning based on neural network has a great potential for realizing real-time GW searches. The main advantages of machine learning are greater computation speed than the traditional algorithms for weak signal extraction and generalization. Presently, the matched-filtering data analysis technique \cite{ref17} is a standard method used for searching GW signals. However, for this technique, the theoretical waveform templates should sufficiently complete and the detected data stream should match with each template. These requirements require a huge amount of computing resources and result in low time effectiveness. For machine learning, once the learning structure is trained well, the detection or prediction results can be obtained rapidly. Generalization is another advantage of machine learning. Both the matched-filtering method and Bayesian model selection \cite{addTaylor,ref20,ref21} for GW searches rely on theoretical template banks; hence, it is possible that we may fail to capture the GW signals that are not included in the template banks. Since machine learning is based on data, it is more likely to explore GW signals beyond the current templates. Although considerable progress is necessary to apply machine learning to real-life pulsar timing array data as no detected GW signal is available as yet. However, it is of interest to develop this technology using the simulated datasets. In this study, we simulated GW signals by using some specific theoretical waveforms to build our training samples. This is just an initial experiment for applying machine learning to real data. In fact, the training sets should be real data (including true events and pure noise); however, no GW event has been detected in the PTA frequency band as yet. Therefore, we used theoretical waveforms to create positive samples. In the future, more GW events will be detected, and related large databases will be built. Then, machine-learning methods can be used for classification and prediction without any theoretical model. Based on simulated datasets, many studies have been conducted on glitch classification \cite{r1,ref18,ref19}, detection, and parameter estimation of GWs in LIGO/Virgo frequency band by using deep-learning algorithms \cite{ref7,ref4,ref5,ref6,r2,addcao,ref26,ref29,ref31}. These studies show that the application of deep learning to search for GWs from binary black hole mergers can provide similar detection sensitivity and computation several orders of magnitude faster than those of matched-filtering, and can be significantly better than the traditional machine-learning techniques. Furthermore, the most important advantage is that deep learning can be extended to search other types of GW signals in addition to the GWs from binary black hole mergers \cite{ref4,ref5}. Meanwhile, the parameter estimation using machine-learning methods in other astronomical observations has also achieved rapid development; for example, convolutional Neural Networks (CNNs) have been used for automated analysis and determination of uncertainties in the parameter estimation of strong gravitational lenses \cite{BNN1,BNN2}; recurrent neural network --Bayesian neural networks (RNN--BNN) have been used to modify dark energy models at higher redshift \cite{BNN3}. Inspired by these studies, we adopted neural networks with different architectures to search the simulated GW signal in PTA data from multiple pulsars and to estimate the chirp mass of the GW source and the distances of pulsars. The results presented in this paper indicate that the CNN is effective for detecting GWs and for noise classification in PTA time series. Such data processing may help detect the GW with the combined SNR $\geq$1.33 for our simulated data (considering the specific 3 pulsars and Gaussian white noise) from SMBBHs with $D_{L}\sim75\text{Mpc}$. Considering the SMBBHs inspiralling with a small eccentricity, the impact of eccentricity on our identification and parameter estimation is also discussed in detail in this paper. Meanwhile, by training the networks with prior information, the feedforward networks (e.g., Elman neural network) obtained are more suitable for parameter estimation. The BNNs can acquire the uncertainty in chirp mass estimation after finding the optimal weight distribution. Hence, our results are more reliable, and the confidence regions are greater.
	
	This paper is organized as follows: In Section II, we describe the detailed process of simulating PTA timing residuals.
	Section III provides the structure of the deep-learning network as a classifier and the corresponding results. In section IV, the mean relative errors of the Elman Neural Network and BNN as adopted for parameter estimation are discussed. Finally, Section V describes some meaningful results and prospects.
	
	\section{Obtaining the simulated data}
	Building datasets is a crucial process in machine learning because it is the original source of information for the computer. Our present understanding of both the gravitational waves and the corresponding noise processes is incomplete. However, as a preliminary study for PTA data analysis, it is important to simulate the data composed of the timing residuals from GW and the simplest noise process (i.e., only stationary Gaussian white noise).
	
	The influence of GW on pulse arrival times (TOAs) can be derived from the GW-induced redshift of a pulse as follows \cite{ref30}:
	\begin{equation}\label{1}
		z(t, \Omega)=\frac{1}{2} \frac{\hat{u}^{a} \hat{u}^{b}}{1+\hat{\Omega} \cdot \hat{u}} \Delta h_{a b}(t, \Omega),
	\end{equation}
	where $\hat{\Omega}$ is a unit vector defining the direction of GW propagation; $\Delta h_{ab}=h_{ab}(t,\hat{\Omega})-h_{ab}(t_{p},\hat{\Omega})$ is the difference in the metric perturbation between the Earth and the pulsar; $t$ and $t_{p}$ are the times when the GW passes the solar system barycenter and the pulsar, respectively. Their relationship is given as $t_{p}=t-d_{p}(1+\hat{\Omega}\cdot\hat{u})$, where $d_{p}$ is the distance to the pulsar, and $\hat{u}$ is a unit vector pointing from the Earth to the pulsar. Then, an offset pulsar TOA (i.e., timing residuals) can be calculated by integrating the GW-induced redshift over the whole observing time as follows:
	\begin{equation}\label{2}
		s(t)=\int_{0}^{t}z(t')dt'.
	\end{equation}
	The timing residuals for an individual GW source can be derived as follows:
	\begin{equation}\label{3}
		s(t,\hat{\Omega})=(F^{+}(\hat{\Omega})\cos2\psi+F^{\times}(\hat{\Omega})\sin2\psi)\Delta s_{+}(t)+(F^{+}(\hat{\Omega})\sin2\psi-F^{\times}(\hat{\Omega})\cos2\psi)\Delta s_{\times}(t),
	\end{equation}
	where $\psi$ is the polarization angle of GW, $F^{+}(\hat{\Omega})$ and $F^{\times}(\hat{\Omega})$ are the antenna pattern response functions that encode
	the response of a given pulsar to a particular GW source \cite{ref2,ref11,ref16}:
	\begin{equation}\label{4}
		\begin{aligned} F^{+}=& \frac{1}{4(1-\cos \theta)}\left\{\left(1+\sin ^{2} \delta\right) \cos ^{2} \delta_{p} \cos \left[2\left(\alpha-\alpha_{p}\right)\right]\right.\\ &-\sin 2 \delta \sin 2 \delta_{p} \cos \left(\alpha-\alpha_{p}\right)+\cos ^{2} \delta\left(2-3 \cos ^{2} \delta_{p}\right) \}, \\ F^{ \times}=& \frac{1}{2(1-\cos \theta)}\left\{\cos \delta \sin 2 \delta_{p} \sin \left(\alpha-\alpha_{p}\right)\right.\\ &-\sin \delta \cos ^{2} \delta_{p} \sin \left[2\left(\alpha-\alpha_{p}\right)\right] \}, \end{aligned}
	\end{equation}
	where $\cos \theta=\cos \delta \cos \delta_{p} \cos \left(\alpha-\alpha_{p}\right)+\sin \delta \sin \delta_{p}$, $\delta$ and $\alpha$ are the declination and right ascension of the GW source, while $\delta_{p}$ and $\alpha_{p}$ are the declination and right ascension of the pulsar. On the other hand, the GW-induced timing residuals should be considered as the comprehensive effects of the Earth and the pulsar's conditions:
	\begin{equation}\label{5}
		\begin{array}{c}{\Delta s_{+, \times}(t)=s_{+, \times}(t)-s_{+, \times}\left(t_{p}\right)}, \\ {t_{p}=t-d_{p}(1-\cos \theta) / c,}\end{array}
	\end{equation}
	where $s_{+,\times}(t)$ is the Earth term, and $s_{+,\times}(t_{p})$ is the pulsar term. According to the Peter--Mathews GW waveforms\cite{ref10}, the pulsar timing residuals induced by eccentric SMBBHs\cite{ref11} are given as follows:
	\begin{equation}\label{6}
		\begin{aligned} s_{+}(t)=& \sum_{n}-\left(1+\cos ^{2} \iota\right)\left[a_{n} \cos (2 \gamma)-b_{n} \sin (2 \gamma)\right] \\ &+\left(1-\cos ^{2} \iota\right) c_{n}, \\ s_{ \times}(t)=& \sum_{n} 2 \cos \iota\left[b_{n} \cos (2 \gamma)+a_{n} \sin (2 \gamma)\right], \end{aligned}
	\end{equation}
where
	\begin{equation}\label{7}
		\begin{aligned} a_{n}=&-\zeta \omega^{-1 / 3}\left[J_{n-2}(n e)-2 e J_{n-1}(n e)+(2 / n) J_{n}(n e)\right.\\ &+2 e J_{n+1}(n e)-J_{n+2}(n e)]\sin [n l(t)], \\ b_{n}=& \zeta \omega^{-1 / 3} \sqrt{1-e^{2}}\left[J_{n-2}(n e)-2 J_{n}(n e)\right. \\ &+J_{n+2}(n e) ] \cos [n l(t)], \\ c_{n}=&(2 / n) \zeta \omega^{-1 / 3} J_{n}(n e) \sin [n l(t)]. \end{aligned}
	\end{equation}
	Here, $s_{+}$ and $s_{\times}$ represent pulsar timing residuals corresponding to the $+$, $\times$ polarization modes of gravitational waves; $J_{n}$ is the Bessel function; $\zeta=(G \mathcal{M})^{5 / 3} / c^{4}D_{L}$, where $\mathcal{M}$ is the chirp mass $\mathcal{M}^{5 / 3}=m_{1}m_{2}(m_{1}+m_{2})^{-1/3}$ ($m_{1}$ and $m_{2}$ are the component masses of binary black holes); $D_{L}$ is the luminosity distance of binary black holes; and $e$ is the eccentricity. Further, $\omega=2 \pi f$; $f$ is the orbital frequency; $l(t)$ is the mean anomaly; $\gamma$ is the initial angle of periastron; and $\iota$ is the inclination angle of the orbit of binary black holes.

	In this study, we chose the following pulsars: PSR J0437$-$4715, PSR J1713$+$0747, PSR J1909-3744. These pulsars have been established as millisecond pulsars with highest long-term timing accuracy\cite{addpulsar, ref2}.
	According to above equations, the simulated GW-induced timing residuals with stationary Gaussian white noise for the timing data of each pulsar can be obtained. All the parameters for the simulation are listed in Table~\ref{para}.

	\begin{table}[ht]

		\newcommand{\tabincell}[2]{\begin{tabular}{@{}#1@{}}#2\end{tabular}{l}}
	    \centering
		\caption{Pulsars and gravitational wave (GW) source parameters for the simulation in this study. Here, $D_{\text{p1}}, D_{\text{p2}}, and D_{\text{p3}}$ are the distances from PSR J1909$-$3744, PSR J1713$+$0747, and PSR J0437$-$4715, respectively. Because of measurement error, the pulsar distances have uncertainties. Sampling frequency: uniform sampling, sampling 393 times in 12 years; noise: gaussian white noise, which is identically distributed $N(0,\sigma^{2})$ ($\sigma=100ns$ for all pulsars).}\label{para}

			\begin{tabular}{cc}
\toprule[2pt]
		
Parameters for Simulation & Values\\
\midrule[1pt]
\multirow{3}{*}{Pulsar (PSR J1909$-$3744)} & {$\alpha_{p1}=5.02~\text{rad}$}  \\
		& {$\delta_{p1}=-0.66~\text{rad}$} \\
        & {${D}_{\text{p1}}$= $ 1.23 $ $\sim$ $ 1.29$ kpc}   \\
        \space\\
\multirow{3}{*}{Pulsar (PSR J1713$+$0747)} & {$\alpha_{p2}=4.51~\text{rad}$}  \\
		& {$\delta_{p2}=0.14~\text{rad}$} \\
         & {${D}_{\text{p2}}$= $ 1.9 $ $\sim$ $ 4.3 $ kpc } \\
         \space\\
\multirow{3}{*}{Pulsar (PSR J0437$-$4715)} & {$\alpha_{p3}=1.21~\text{rad}$}  \\
		& {$\delta_{p3}=-0.82~\text{rad}$} \\
        & {${D}_{\text{p3}}$= $ 0.1565 $ $\sim$ $0.1571 $ kpc} \\
        \space\\
            	
\multirow{2}{*}{GW source} & {$\alpha=3.74~\text{rad} $} \\
		& {$\delta= 0.44~\text{rad}$}\\
\space\\

				Luminosity distance of binary black holes (\text{Mpc}) & 75\\
				Orbital frequency of binary black holes (\text{Hz}) & ~~~~$2.15\times 10^{-9}$ $\sim$ $2.43\times 10^{-9}$\\
				Eccentricity & $0$\\
				Chirp mass ($M_{\odot}$) & $10^{8}$ $\sim$ $10^{9}$ \\
				Orbital inclination of binary black holes (rad) & $0.36 $ $\sim$  $0.48$\\
				Polarization of Gravitational Wave & $ 0 $ $\sim$ $ \pi $\\
				Initial pericentric angle (rad) & $ 3.23 $ $\sim$ $3.84$\\
				Initial phase & $ 0.72 $ $\sim$ $0.83$ \\
				\bottomrule[2pt]
			\end{tabular}
	\end{table}
	
\begin{figure*}[!htb]
\centering
	\includegraphics[width=0.7\textwidth]{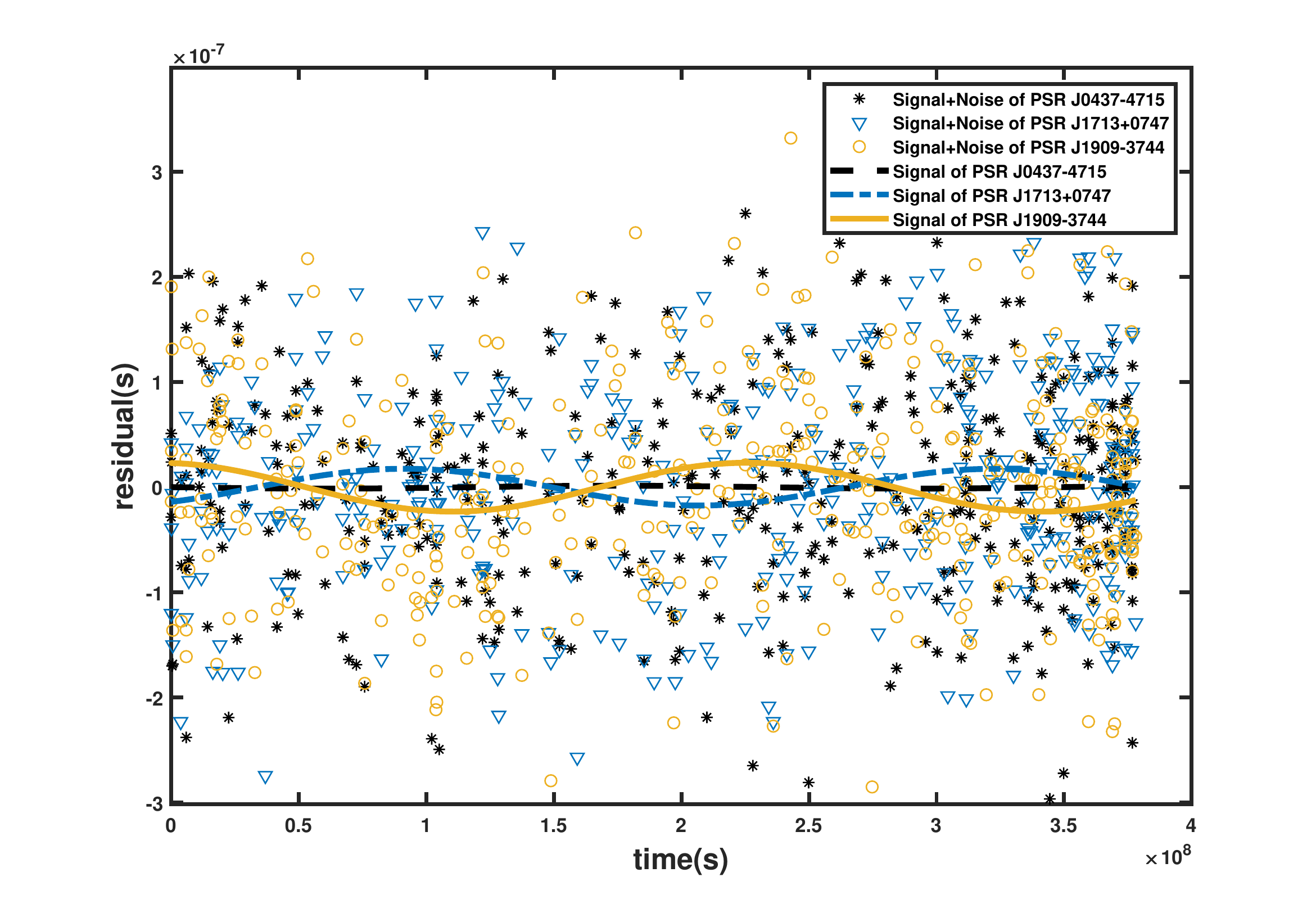}

 \caption{Simulated time-domain data of the PTA as one sample in our training sets, which is in $3\times393$ dimension; the black star points and dashed line represent simulated timing residuals$+$noise and expected timing residuals from GW, respectively, for PSR J0437$-$4715.The yellow circles and solid line represent simulated timing residuals$+$noise and expected timing residuals from GW, respectively, for PSR J1909-3744. The blue triangles and dotted line represent simulated timing residuals$+$noise and expected timing residuals from GW, respectively, for PSR J1713+0747.}
    \label{figmassh}
 \end{figure*}

\begin{figure*}[!htb]
\centering
	\includegraphics[width=1.05\textwidth]{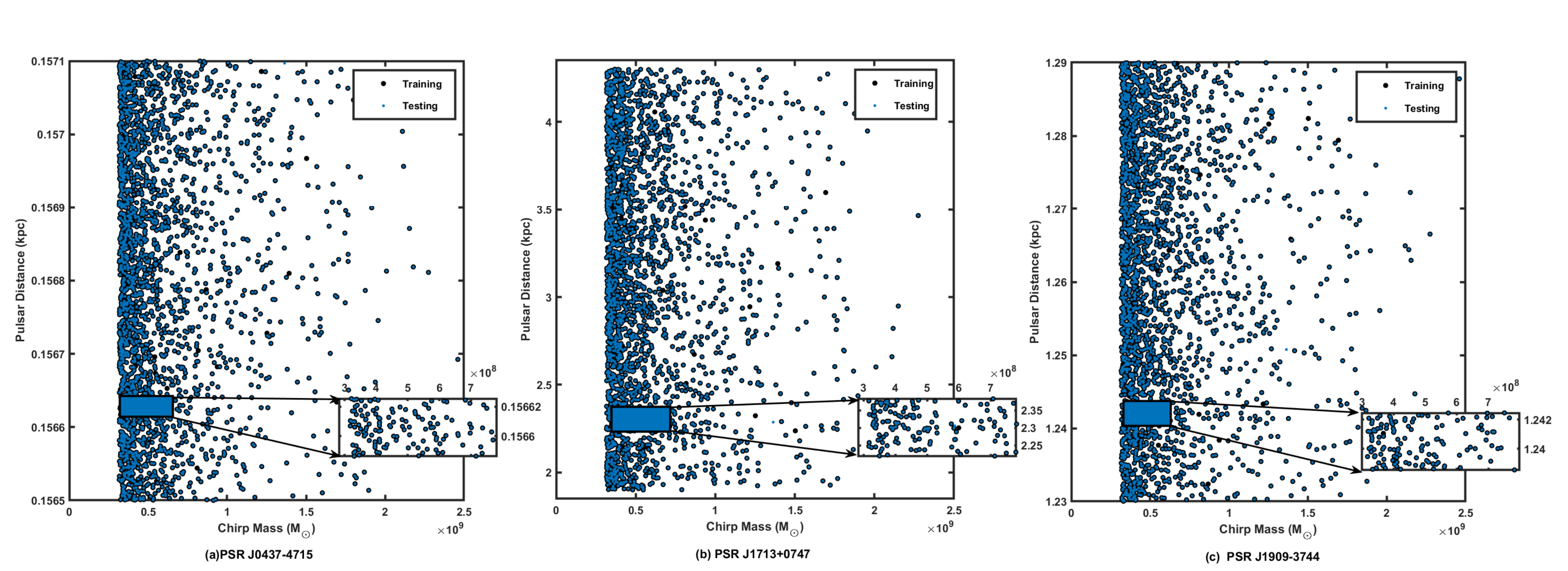}

    \caption{Distribution of simulated data sets in the chirp mass bank. The blue points represent the training sets, and the black points represent the testing sets. In the inset, the distribution of chirp mass intervals is magnified. Testing sets and training sets are distributed similarly, but they do not overlap.}
 \label{distribution}
\end{figure*}

	Assuming the noise in each sample is additive, the data of multiplied pulsar timing residuals can be written as follows:
	\begin{equation}\label{8}
		D_{i}=(d_{1i};d_{2i};d_{3i})~~~~~~~~~~~d_{ki}=s_{ki}+n_{ki}, (k=1,2,3)
	\end{equation}
	where $s_{ki}=s(t_{i},\boldsymbol{\lambda})$ and $n_{ki}=n(t_{i})$ are the GW-induced timing residuals in Eq.(\ref{3}), and the noise of the $k^{\text{th}}$ pulsar, $\boldsymbol{\lambda}$, is the source parameter. In this study, we generated 300,000 samples with different sources and pulsar parameters. Then, we randomly chose 15,000 for training (containing multiple eccentricities $e=0, 0.05, 0.1$) and another 10,000 (also including $e=0, 0.05, 0.1$) for testing. Each sample is similar to that shown in Fig.\ref{figmassh}, where both pulsar and Earth terms are considered. Moreover, their frequencies are supposed to be identical because the frequency will not evolve when the mass of the black hole $<\sim10^{9}$ M$_{\odot}$\cite{ref1}. The typical frequency evolution time for the SMBBH in that mass range is more than 10$^{6}$ years, while the time range considered in the simulations is $\sim$10 years. To investigate the ability of our neural network for different GW strengths, the training and testing sets were sorted into several subsets with different signal to noise ratios (SNRs). For each subset, the corresponding SNR of individual pulsar is given as follows:
	\begin{equation}\label{9}
		\text{SNR}=\sqrt{(s_{i}|s_{i})},
	\end{equation}
	where $(s_{i}|s_{i})=\sum_{i}s_{i}\cdot s_{i}/\sigma^{2}$; $\sigma^{2}$ is the mean square of the residuals from the pulsar, which is also equal to the power of noise. In our case, $\sigma=100\text{ns}$, and sampling frequency is 393 times in 12 years. The combined SNR can be defined as
 \begin{equation}\label{10}
	\text{SNR}_{com}=\sqrt{\text{SNR}_{1}^{2}+\text{SNR}_{2}^{2}+\text{SNR}_{3}^{2}},
	\end{equation}	
where $\text{SNR}_{j}$ represents the $j^{\text{th}}$ pulsar. At the same time, we ensured there was no overlap between the training and testing sets, as shown in Fig~\ref{distribution}.

\section{Machine learning for identifying TOA variations from individual inspiralling supermassive binary black holes}
 The neural network model we used here is composed of convolutional layers and fully connected neural networks, which have been used in previous studies to search GW signals and for parameter estimation in LIGO/Virgo frequency band\cite{ref4,ref5,ref6,ref7,r2,addcao,ref12,ref13,ref14}, and employed in the glitch classification\cite{r1} and signal denoising \cite{wen56,wen57,wen58,wen59}. More importantly, this model has also been used to detect GW waveforms with higher-order multipoles from eccentric binary mergers \cite{wen60}. Motivated by these studies, we chose the CNN as the basic model for GW signal detection in the PTA frequency band.

 The architecture of our CNN as a classifier is listed in Table~\ref{fig:classifier}. The main part of a typical CNN is the convolutional layer, which contains a set of neurons that share their weights and execute convolution operation output from the upper layer and a fixed size kernel \cite{deeplearning}. Then, the pooling layer pools the result from the upper convolution layer, usually choosing the maximum value in each convolution kernel to further reduce the computational cost and make the network more sensitive to data features \cite{maxpool}. After the pooling layer, the active function is used to realize nonlinear fitting. Here, $``$Ramp$"$ represents Relu (Rectified Linear Unit) function as the active function. It was chosen owing to its fast convergence\cite{relu,fullconnected}. With the deepening of the convolution layer, an increasing amount of abstract information will be extracted. The flattened layer is used to $``$flatten$"$ the input, that is, to unilateralize the multi-dimensional input. It is often used in the transition from the convolution layer to the full connection layer\cite{fullconnected}. Some full connection layers marked as $``$DotPlusLayer$"$ play the role of $``$Classifier$"$ in the entire convolution neural network and output multiple scalars in the expected dimension for further classification. The $``$BatchNormalizationLayer$"$ inserted between convolution layer and full connection layer is used to change data with a nonstandard distribution in the training process back into the standard normal distribution with zero mean value and one variance\cite{batchnorm}. Finally, the $``$SoftmaxLayer$"$ is often used in the last layer of neural network as the output layer for binary or multiclassification to map multiple scalars to a probability distribution\cite{softmax}.

A CNN has several hyperparameters, such as the number of neurons in the convolutional layers, kernel size, and dilation factor of the convolutional layers. In this study, most of the hyperparameters were fixed through manual adjusting, and their optimal values are listed in the caption of Table~\ref{fig:classifier}. The training process begins with samples in larger SNR intervals, and then gradually switches to the ones in the lower SNR intervals. Hence, the network can extract hidden information more easily and accurately \cite{ref7}.

 \begin{table}[ht]
\centering
\caption{Structure of the neural network used as a classifier. The number of neurons in the convolutional layers is 32, 16, 16, and 32. The kernel sizes are 1$\times$3 for all the convolutional layers and 1$\times$2 for all the pooling layers. The stride is set as 1 for all of the convolutional layers and pooling layers. The dilation factor of the convolutional layers is set as 1, and the padding size of the pooling layer is zero. The function of pooling is max. The active function is Relu.}\label{fig:classifier}
\begin{tabular}{lc}

\toprule[2pt]
Input&Matrix(1$\times$1179)\\
\midrule[1pt]
1~ReshapeLayer~~~~~~&3-tensor(size:1$\times$1$\times$1179)\\
2~ConvolutionLayer&3-tensor(size:32$\times$1$\times$1177)\\
3~Ramp~~~~~~~&3-tensor(size:32$\times$1$\times$1177)\\
4~PoolingLayer~~~~~~~~~&3-tensor(size:32$\times$1$\times$589)\\
5~ConvolutionLayer&3-tensor(size:16$\times$1$\times$587)\\
6~Ramp~~~~~~~&3-tensor(size:16$\times$1$\times$587)\\
7~PoolingLayer~~~~~~~~~~~~~&3-tensor(size:16$\times$1$\times$294)\\
8~ConvolutionLayer&3-tensor(size:16$\times$1$\times$292)\\
9~Ramp~~~~~~~&3-tensor(size:16$\times$1$\times$292)\\
10~PoolingLayer~~~~~~~~~&3-tensor(size:16$\times$1$\times$146)\\
11~ConvolutionLayer&3-tensor(size:32$\times$1$\times$144)\\
12~Ramp~~~~~~~&3-tensor(size:32$\times$1$\times$144)\\
13~PoolingLayer~~~~~~~~~&3-tensor(size:32$\times$1$\times$72)\\
14~BatchNormalizationLayer~~~~~~~&3-tensor(size:32$\times$1$\times$72)\\
15~FlattenLayer~~~~~~~&vector(size: 2304)\\
16~LinearLayer~~~~~~~~&vector(size: 16)\\
17~DropoutLayer~~~~~~~~&vector(size: 16)\\
18~LinearLayer~~~~~~~~~&vector(size:2)\\
19~SoftmaxLayer~~~~~~&vector(size: 2)\\
Output~~~~~~~~~~&class\\
\bottomrule[2pt]
\end{tabular}
\end{table}

In this process, our network was initialized by the $``$Xavier$"$ method. Given that the initial learning rate and L2 regularization coefficient are 5$\times 10^{-4}$ and 0.1, respectively, the ADAM algorithm, which is an optimization algorithm based on the traditional stochastic gradient descent algorithm, is chosen. It can iteratively update the weights of neural networks according to the training data. Furthermore, the learning rate can be adjusted automatically based on the first moment mean\cite{ref15}.

In supervised binary classification, the datasets including GW signal are labeled as Ture; otherwise, they are labeled as False. The last layer is chosen as SoftmaxLayer, which indicates that the softmax cross entropy loss is used.

\begin{figure*}[!htb]
\centering
	\includegraphics[width=1.05\textwidth]{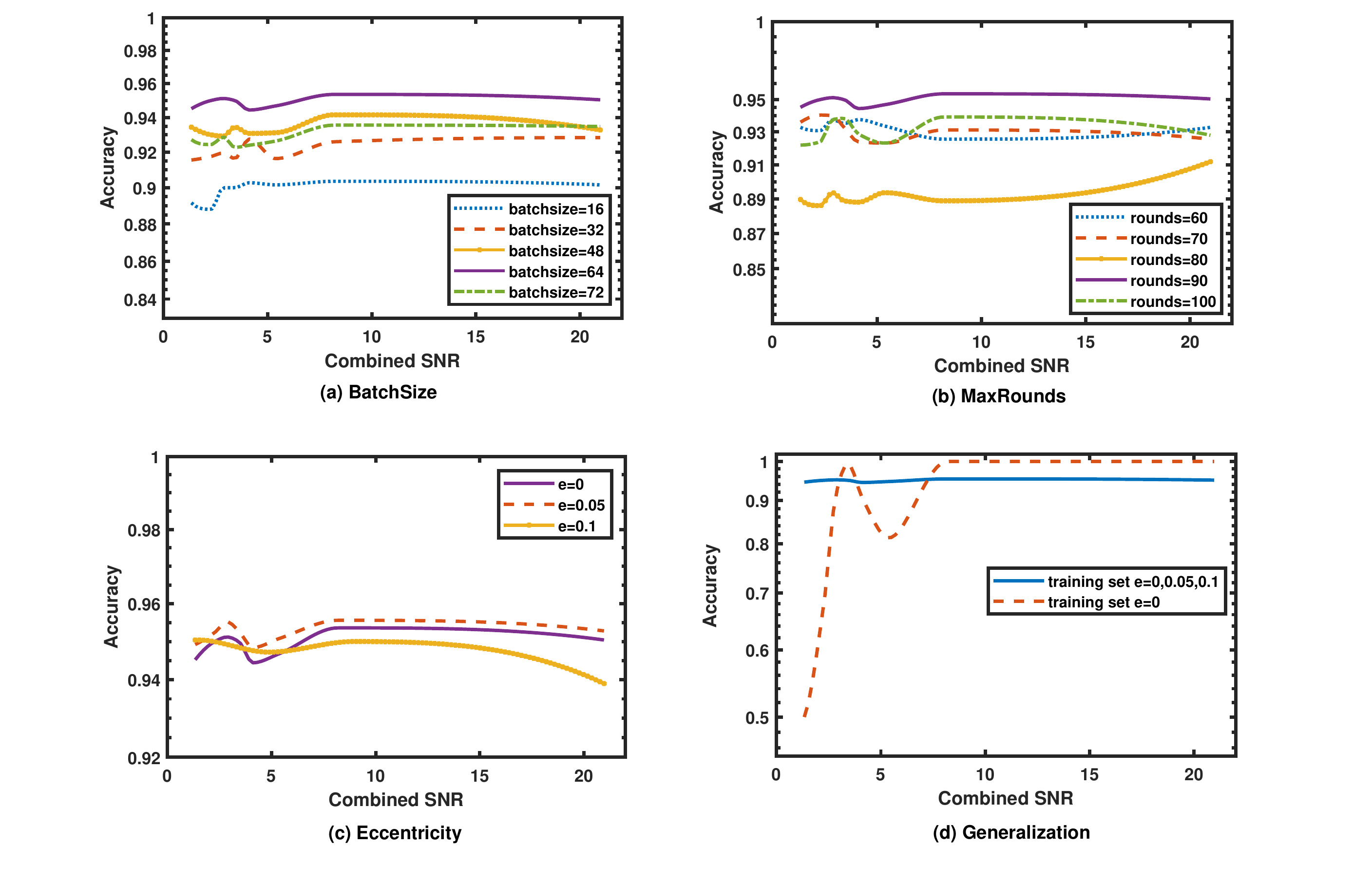}
    \caption{(a) Test accuracies versus combined SNR with different batch sizes (here, maxrounds was set to be 100, and the eccentricity of the testing datasets is 0); (b) the test accuracies versus combined SNR with different rounds (here, batch size was set to be 64, and the eccentricity of the testing datasets was 0); (c) the test accuracies vary with combined SNR for the testing data sets having different eccentricities; (d) the accuracies verse combined SNR with different training sets (one consists only of samples with $e=0$, and the other, samples with $e=0, 0.05, and 0.1$).}
    \label{snraccu}

\end{figure*}

On Mathematica 11.1 platform, training the CNN network takes approximately 0.4 h with NVIDIA GPU (Gtx1080ti, 11 GB RAM). Meanwhile, MaxTrainingRounds and Batchsize both have great impacts on the performance of the network. The test accuracies for identifying TOAs from GW with different Batchsize and MaxTrainingRounds are illustrated in Fig~\ref{snraccu}(a), (b). The results show that on the whole, with the combined SNR increasing, the accuracy is enhanced, while oscillation is observed for the lower SNR interval, and the optimal Batchsize and Maxtrainingrounds are 64 and 90, respectively. In this case, our CNN detection accuracy can reach more than 94\% when SNR$\geq1.33$. Furthermore, the accuracy of each eccentricity varies with the combined SNR as shown in Fig~\ref{snraccu}(c), and the network can have a high test accuracy. The accuracy seems to decrease for SNR$>$15 when e=0.1 because there are fewer training samples for e$=0.1$ in such SNR intervals. When the combined SNR exceeds 8, the network can still recognize the samples with non-zero eccentricity even if all the training sets are zero (ref. Fig~\ref{snraccu}(d)). Therefore, the generalization of our network is guaranteed on the whole. Similarly, the number of training samples around combined SNR$\sim$6 is less than the samples in the nearby SNR intervals, resulting in a sharp drop in accuracy.

\section{Machine learning based on neural networks for source and pulsars parameter estimation}

\begin{figure*}[!htb]
\centering
	\includegraphics[width=0.8\textwidth]{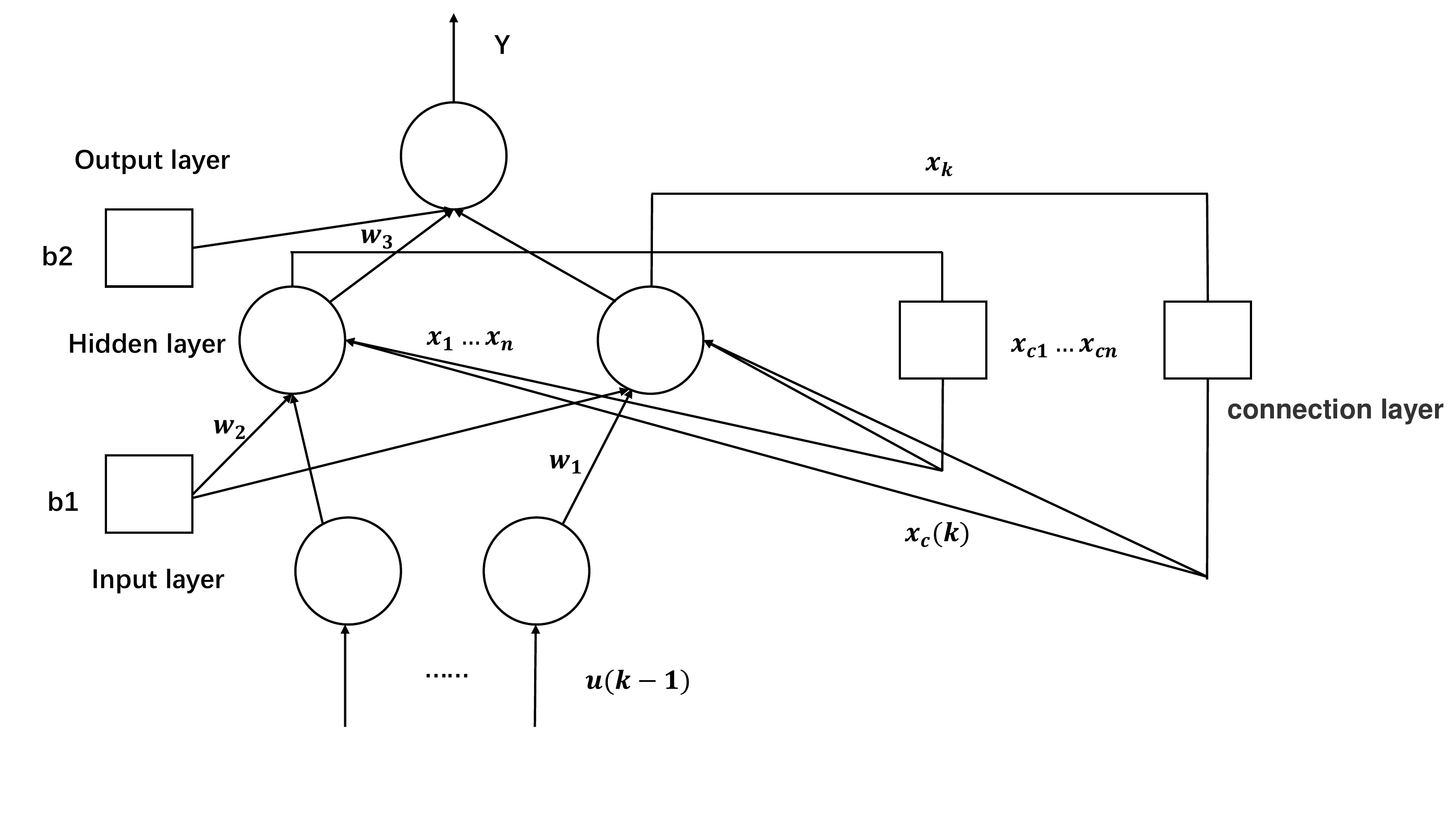}
    \caption{$Y$ is the m-dimension output node-vector, $x$ is the N-dimension intermediate node elements vector, $u$ is the R-dimension input vector, $x_{c}$ is the N-dimension feedback vector, $w_{3}$ is the connection weight from the middle layer to the output layer, and $w_{2}$ is the connection weight from the input layer to the middle layer, $w_{1}$ is the connection weight between the connection layer and the middle layer \cite{elman}.}
    \label{struelman}

\end{figure*}

 For GW sources, the degeneracy in luminosity distance and chirp mass leads to concerns regarding the estimation of chirp mass $\mathcal{M}$, given the luminosity distance $D_{L}$. For multiple pulsars, their luminosity distance $D_{\text{p}i} (i=1,2,3)$ can be estimated. Considering the time dependence for each data point, we find that the Elman Neural Network provides much better results in estimating the source parameter than the CNN \cite{elman1}. The RNN has achieved much success and has wide applications in natural language processing \cite{ref23,ref24}. There are similarities between the GW signals and speech recognition (e.g., the data of the latter time depends on the previous data points). Therefore, the associative memory function of the network is better and more stable. The Elman neural network is a typical local regression network. It can be regarded as a recurrent neural network with a local memory unit and local feedback connection. Compared with other traditional RNNs such as, GRU and LSTMs, the Elman neural network has a simpler structure to realize the basic properties such as sensitivity to the data of historical states and strong ability to process dynamic information, which are similar to the properties of the GW sequence. Elman neural network generally has four layers, namely,
input layer, hidden layer, connection layer, and output layer. Input layer, the output layer and the hidden layer are similar to the structure of feedforward network. The input layer elements only play the role of signal transmission, while the output layer elements work in linear weighting. The transfer function of the hidden layer element can be linear or nonlinear. The connection layer is used to record the output values of the hidden layer elements at the previous moment and return them to the network, which is a delay operator actually \cite{elman}. The structure of Elman Neural Network is shown in Fig~\ref{struelman}.

\begin{table}[ht]
	\centering
	\caption{The hyperparameters of applying Elman Neural Network to estimate chirp mass $\mathcal{M}$ of GW source and luminosity distance $D_{pi}$ of the pulsars. The dimension of input is 1$\times$1179 (the dataset of each pulsar is in 1 $\times$393 dimension), the training function is $``$ traingdx $"$ (Gradient descent method with momentum and adaptive learning rate), Max-fail is the maximum number of validation failures, which means the training process should be stopped, once the amount of invalid training on validation datasets is more than max-fail.}
	\label{paraelman}

	\begin{tabular}{cc}
		\toprule[2pt]
		Hyper-parameter & Value\\
		\midrule[1pt]
		Neuron Number & $15$ ($\mathcal{M}$ estimation) and $20$ ($D_{p}$ estimation)\\
		Learning Rate &$10^{-4}$\\
		Maximum of Iterations &   ~~~$1800$ ($\mathcal{M}$ estimation) and $3000$ ($D_{p}$ estimation)\\
		Error Tolerance &  $10^{-5}$\\
		Max-fail & $100$\\
		Loss Function & Mean Squared Error (MSE)\\
		\bottomrule[2pt]
		
	\end{tabular}
\end{table}

\begin{figure*}[!htb]
\centering
	\includegraphics[width=0.6\textwidth]{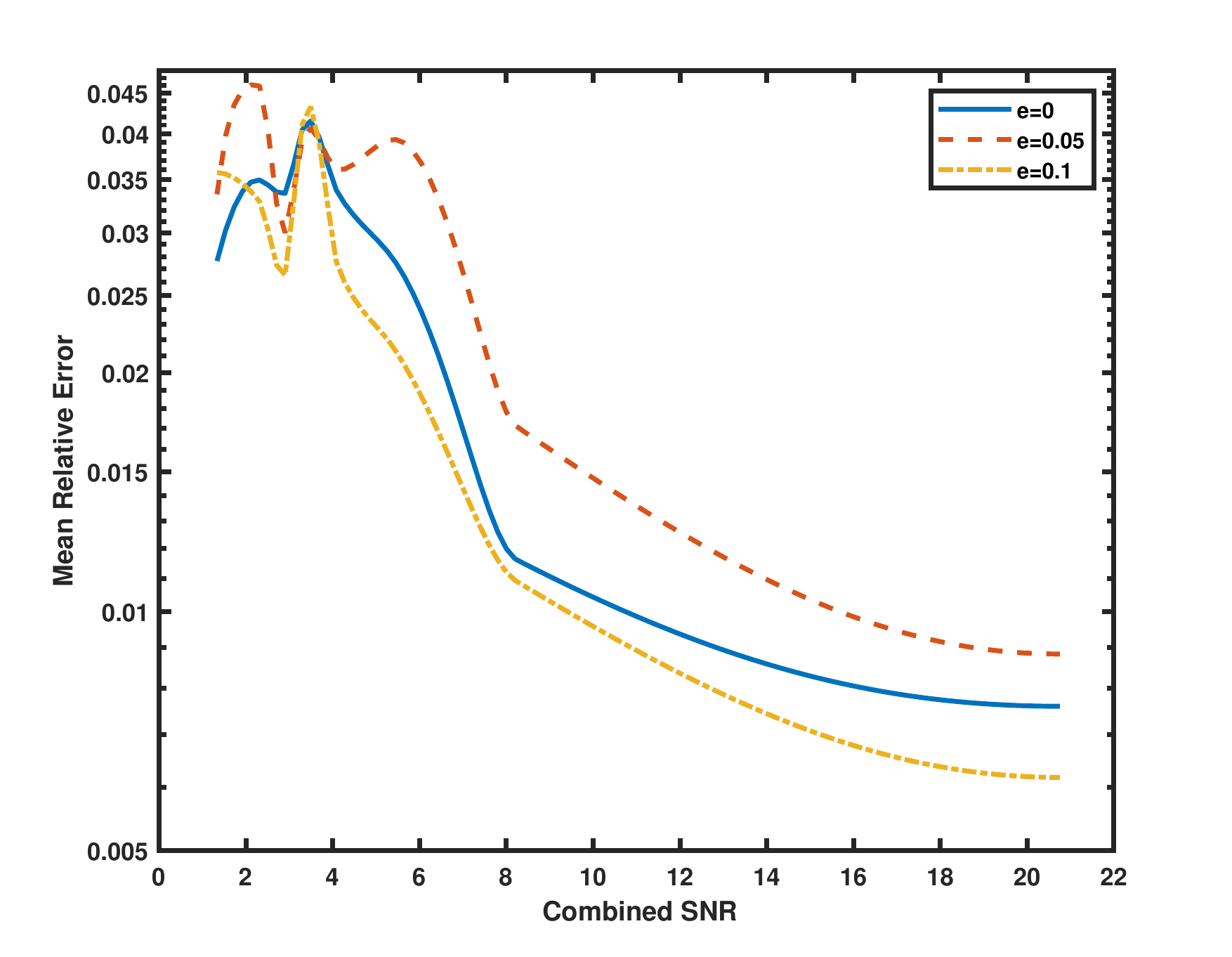}
    \caption{The mean relative error of chirp mass $\mathcal{M}$ varies with combined SNR; The blue, red, and yellow curves refer to the cases of e$=0, 0.05, 0.1$, respectively.}
    \label{paramass}

\end{figure*}

\begin{figure*}[!htb]
\centering
	\includegraphics[width=1.05\textwidth]{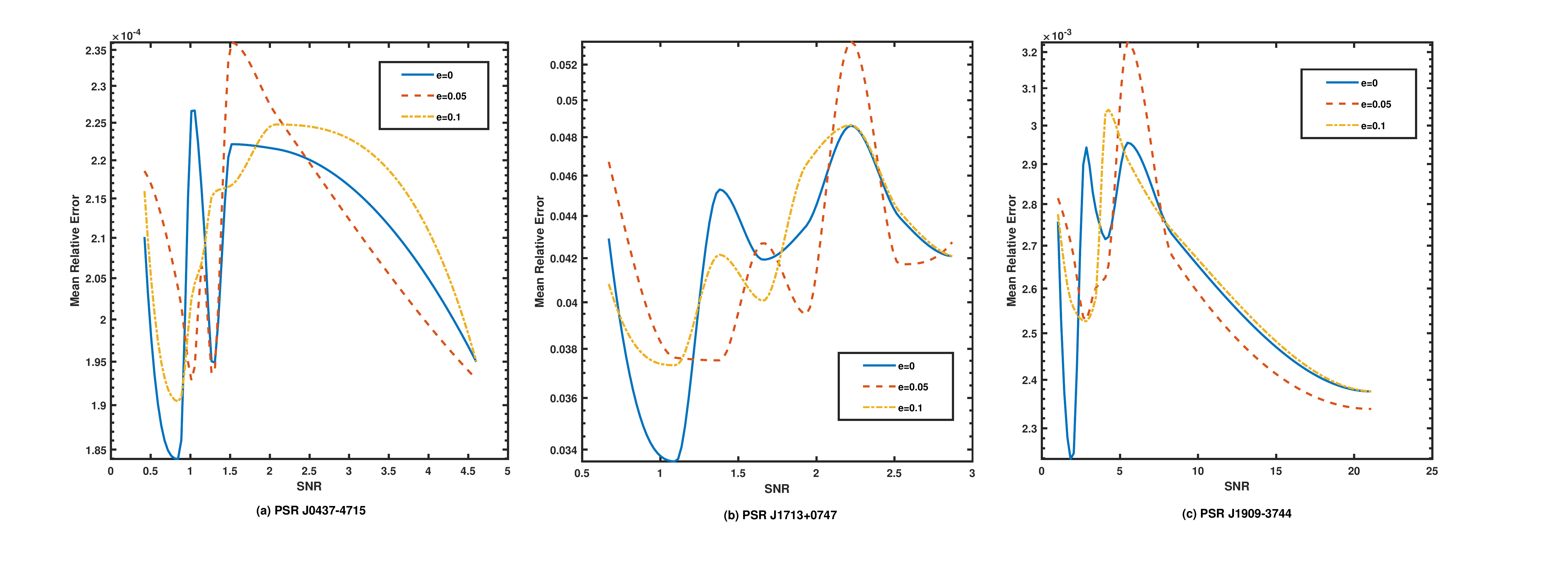}
    \caption{(a) The mean relative errors of $D_{p}$ estimation vary with the SNR of PSR J$0437$-$4715$; (b) The mean relative errors of $D_{p}$ estimation vary with the SNR of PSR J$1713$+$0747$; (c) The mean relative errors of $D_{p}$ vary with the SNR of PSR J$1909$-$3744$. The cases with $e=0, 0.05, 0.1$ are shown in blue, red and yellow. The SNR values in the x$-$axis for the three pulsars are all the individual SNRs which are obtained from Eq.(9)}
    \label{parah}

\end{figure*}

Under the framework of Matlab 2018b, training the Elman neural network with different hyperparameters repeatedly, it took about 15 seconds for $\mathcal{M}$ estimation and 25 seconds for $D_{p}$ estimation with GPU each time. The optimal values of some hyperparameters are listed in Table~\ref{paraelman}. Using the mean relative error to measure the performance of our neural network, the corresponding results of the abovementioned two parameters are shown in Fig~\ref{paramass} and Fig~\ref{parah}. For $\mathcal{M}$ estimation, when combined SNR $>$5, the mean relative error is less than 3.7$\%$ and decreases significantly with combined SNR increasing; For $D_{p}$ estimation, it can be found that the mean relative error of PSR J1713+0747 (3.4$\%$ $\sim$ 5.4$\%$) is larger than the errors from other two pulsars (both less than 1$\%$) due to its farther distance and stronger noise, which means in the lower SNR intervals, to a certain extent our network as a predictor would perform worse because of the difficulty in recognizing the characteristic information of the signal and make estimation; In the higher SNR intervals (SNR$>$2.5 for PSR J0437$-$4715 and SNR$>$5 for PSR J1909$-$3744), the mean square errors decrease significantly because the network is easier to recognize the properties with high SNR. Furthermore, the eccentricity would certainly lead to some discrepancies. Totally, the network can well recognize the features of the signal with higher SNR, and the estimation result would get better with the enhancement of the SNR. In the literature \cite{addTaylor}, referring to Fig.4, we can find that the relative error of chirp mass is more than 20$\%$ with SNR=8.

As an important issue of GW observation, the uncertainty of source parameter estimation cannot be calculated from the network using point-estimation on weights, such as CNN, and RNN. Therefore, we choose the Bayesian neural network (BNN), which can predict the distribution of the output using the prior distribution on network's weights and biases, to measure the uncertainty of estimation on objective source parameter \cite{ref25,refbay}.

Using Bayesian approach to estimate the source parameter requires the neural networks to offer a probabilistic value of output, so the weights of these neural networks should have distributions instead of having deterministic values. In other words, our purpose is to calculate the probabilistic distribution of each output value with a new testing input by integrating over all possible weights, which can be expressed as
\begin{equation}
p(y^{*}|x^{*}, X, Y)=\int p(y^{*}|x^{*},\omega)p(\omega|X,Y)d\omega,
\end{equation}
where $X={x_{1},...,x_{N}}$ is the total sample of training input ($x_{i}$ is a 1$\times$393 time-series as Fig.\ref{figmassh} (a)), and $Y={y_{1},...,y_{N}}$ is the corresponding label. The posterior $p(\omega|X,Y)$ is usually approximated by a variational distribution $q(\omega)$. The optimizing process makes $q(\omega)$ as close as possible to the true distribution by minimizing their Kullback-Leibler (KL) distance, which measures the divergence between two distributions. Generally, we chose $q(\omega)$ to be a Gaussian distribution with $\sigma_{\omega}^{2}$ variance and $\mu_{\omega}$ mean value, which can be considered as the prior information. On the other hand, $p(y^{*}|x^{*},\omega)$ means given the weights and input $x^{*}$ the probability of outputting $y^{*}$, in practice which could be chosen as
\begin{equation}
p(y^{*}|x^{*},\omega)=\frac{1}{\sqrt{2\pi}\sigma}\exp[(y^{*}-y)^{T}\sigma^{-1}(y^{*}-y))],
\end{equation}
where $\sigma$ is the prior standard deviation of $y^{*}$, which can be set as the prior standard deviation of weights (i.e., $\sigma=\sigma_{\omega}$).
Then the neural networks must find the optimal $\sigma_{\omega}$ and $\mu_{\omega}$ through repeatedly training the networks with sampled weights from $q(\omega)$.

\begin{figure*}[!htb]
\centering
	\includegraphics[width=1.05\textwidth]{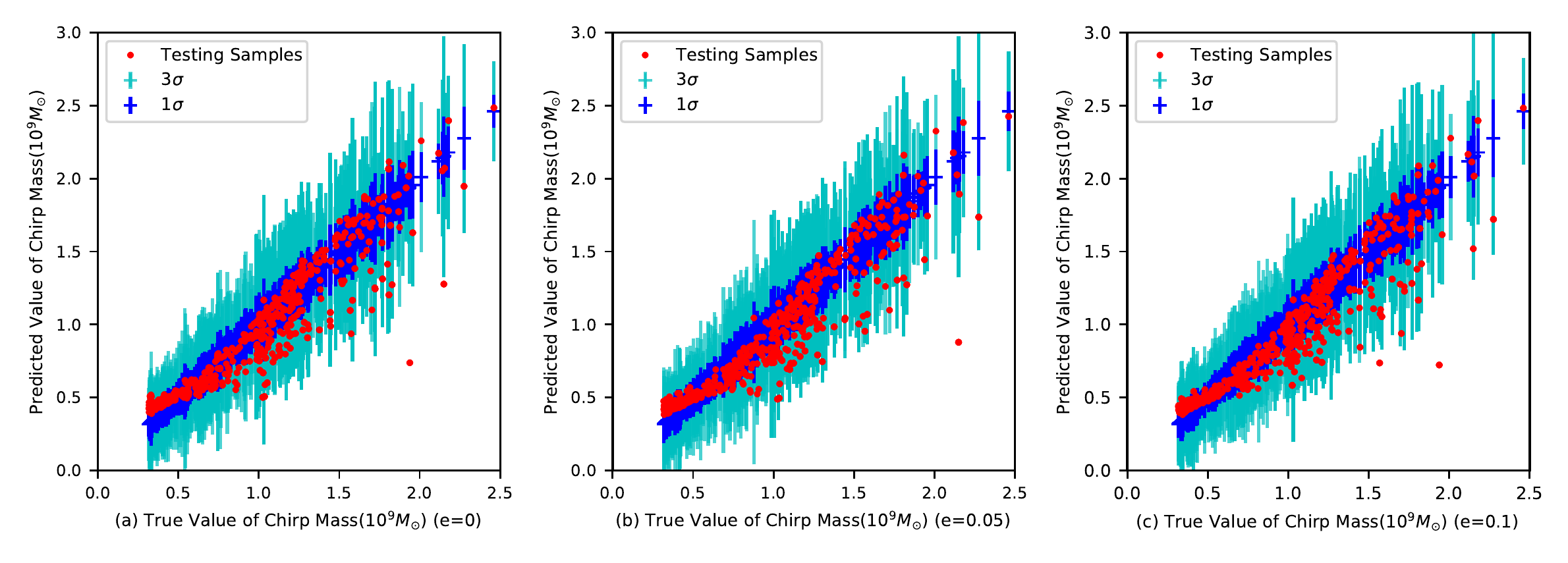}

    \caption{The uncertainty of chirp mass estimation with eccentricity =0, 0.05, 0.1 in subgraph (a), (b), (c), respectively. The uncertainty contours of 1$\sigma$ and 3$\sigma$ are in blue and light blue, respectively; The black points are the testing samples with true values (x-axis) and predicted values (y-axis).}
 \label{uncert}
\end{figure*}

The structure and hyper-parameters of our BNN are listed in Table \ref{bnn} and \ref{paraBayesian}, respectively. In our case, $X$ is a $15000\times1179$ (each of the pulsars has a dataset in 1$\times$393 dimension) matrix, $15000$ indicates the number of samples used in training. $Y$ labels the corresponding chirp mass of $X$, which is $15000\times1$ accordingly. $x^{*}, y^{*}$ represent the testing input and the corresponding expected chirp mass, respectively. Based on tensor flow on python 3.7, the training and testing processes are realized. Fig.\ref{uncert} shows the estimated chirp mass against the true value of this parameter for 1200 testing samples. We can see that all the predicted values are within a normal distribution. Moreover, it shows BNN is sensitive to the chirp mass around $5\times10^{8}M_{\odot}$, and the mean relative error for all the testing dataset is $13.6\%$.

\begin{table}[ht]
\centering
\caption{The structure of the neural network used to combine with Bayesian. A Gaussian distribution is generated, which is generated to obtain observable data, and then three full connection layers are included where neural numbers are 10, 30, and 20, with activation functions are all set to be Relu. Then, the inference is realized by MCMC, which leads to the estimated distribution with mean value $\mu$ and standard deviation $\sigma$ of predicted parameter.}\label{bnn}
\begin{tabular}{lc}

\toprule[2pt]
Structure&Parameter\\
\midrule[1pt]
1~Generative model~~~~~~~~&Gaussian distribution\\
2~Full Connected Layer~~~~~~&vector(size:10)\\
3~Relu~~~~~~~&vector(size:10)\\
4~Full Connected Layer&vector(size:30)\\
5~Relu~~~~~~~&vector(size:30)\\
6~Full Connected Layer~~~~~~~&vector(size:20)\\
7~Relu~~~~~~~&vector(size:20)\\
8~Inference method~~~~&MCMC\\
Output~~~~~~~~~~&Predictive Value $\&$ ($\mu$,$\sigma$)\\
\bottomrule[2pt]
\end{tabular}
\end{table}

\begin{table}[h]
\centering
\caption{The hyper-parameters of Bayesian Neural Network (BNN) used for chirp mass estimation. The dimension of input is 1$\times$1179 (each of the pulsars has a data vector of 1 $\times$393); Given the Gaussian Distribution as prior distribution, the mean square error is set as loss function.}
\label{paraBayesian}
\begin{tabular}{cc}
\toprule[2pt]
Hyper Parameter & Value\\
\midrule[1pt]
Epoch & $20$\\
BatchSize &$10$\\
Learning Rate &   $10^{-6}$\\
Prior Distribution & Gaussian Distribution\\
Active Function & $\text{Relu}$\\
Keep probability of dropout layer & 0.97\\
\bottomrule[2pt]

\end{tabular}
\end{table}

\section{Conclusion}

 We have investigated the effect of machine learning based on neural networks for PTA GW searches and corresponding source parameter estimation. According to the features of GW from individual SMBBHs, the simulated PTA time residuals are generated to train our neural networks. By adjusting the hyperparameters of the neural networks, the CNN, RNN, and BNN with optimal structure are obtained. Although this process takes several hours, the instant classification and parameter estimation with a specific SNR for testing samples can be achieved. From these results, we can determine the fundamental relation between the classification accuracy of the CNN and SNR of the data sets. Furthermore, we adopt a kind of RNN for parameter estimation and obtain much better results than other machine learning methods, which indicates the RNN indeed can extract more information in time series data sets. By using BNN, the error bars of parameter prediction can be obtained, which provides important information for observation.

The major advantage of using machine learning to GW searches is that classification or prediction from a neural network is time-saving once the neural network has been trained well. Meanwhile, note that some factors that must exist in real data are not considered in our simulated data, such as the noise models for each pulsar (the noise process in timing residuals is time-correlated), the possible GW background (GWB) influence and so on. Moreover, for machine learning itself, its performance and generalization are limited severely by the data quality, while the PTA data generally are very noisy. So, there is still a long way to implement these techniques in real data, but it provides some feasible robust approaches for future real PTA data analysis.

\section*{\bf Acknowledgements}
This work was supported by the National Natural Science Foundation of China (Grant Nos. 11873001, 11725313, 11690024), the Natural Science Foundation of Chongqing (Grant No. cstc2018jcyjAX0767), National Key R\&D Program of China No. 2017YFA0402600, by the CAS International Partnership Program No.114A11KYSB20160008, by the CAS Strategic Priority Research Program No. XDB23000000.



\begin{thebibliography}{nbound}
\bibitem{ref33} Liu X J, Pulsar Timing Arrays and the Detection of Gravtational-wave Background. 2016.
\bibitem{ref34} Zhu X J, Detection of Gravitational Waves by Pulsar Timing Array, 2016(2):10-14.
\bibitem{ref1} Zhu X J, Hobbs G, Wen L, et al. An all-sky search for continuous gravitational waves in the Parkes Pulsar Timing Array data set. Monthly Notices of the Royal Astronomical Society, 2014, 444(4): 3709-3720.
\bibitem{ref9} Lee K J, et al. Gravitational wave astronomy of single sources with a pulsar timing array. Monthly Notices of the Royal Astronomical Society 414.4 (2011): 3251-3264.
\bibitem{ref2} Feng Y, Li D, Li Y R, et al. Constraints on individual supermassive binary black holes using observations of PSR J1909-3744. arXiv:1907.03460, 2019.
\bibitem{ref28} Taylor S R, Vallisneri M, Ellis J A, et al. Are we there yet? Time to detection of nanohertz gravitational waves based on pulsar-timing array limits. The Astrophysical Journal Letters, 2016, 819(1): L6.
\bibitem{ref3} Ellis J A, Jenet F A, McLaughlin M A. Practical methods for continuous gravitational wave detection using pulsar timing data. The Astrophysical Journal, 2012, 753(2): 96.
\bibitem{ref17} Nakano H, Narikawa T, Oohara K, et al. Comparison of various methods to extract ringdown frequency from gravitational wave data. Physical Review D, 2019, 99(12): 124032.

\bibitem{addTaylor} Taylor S, Ellis J, and Gair J. Accelerated Bayesian model-selection and parameter-estimation in continuous gravitational-wave searches with pulsar-timing arrays, Phys. Rev. D, 2014, 90, 104028.
\bibitem{ref20} Vigeland S J, Islo K, Taylor S R, et al. Noise-marginalized optimal statistic: A robust hybrid frequentist-Bayesian statistic for the stochastic gravitational-wave background in pulsar timing arrays. Physical Review D, 2018, 98(4): 044003.
\bibitem{ref21} Ellis J A, Cornish N J. Transdimensional Bayesian approach to pulsar timing noise analysis. Physical Review D, 2016, 93(8): 084048.
\bibitem{r1} George D, Shen H, Huerta E A. Classification and unsupervised clustering of LIGO data with Deep Transfer Learning, Phys. Rev. D, 2018, 97, 101501.
\bibitem{ref18} Razzano M, Cuoco E. Image-based deep learning for classification of noise transients in gravitational wave detectors. Classical and Quantum Gravity, 2018, 35(9): 095016.
\bibitem{ref19} Huerta E A, George D, Zhao Z, et al. Real-time regression analysis with deep convolutional neural networks. arXiv:1805.02716, 2018.
\bibitem{ref4} George D, Huerta E A. Deep neural networks to enable real-time multimessenger astrophysics. Phys. Rev. D, 2018, 97(4): 044039.
\bibitem{ref5} George D, Huerta E A. Deep Learning for real-time gravitational wave detection and parameter estimation: Results with Advanced LIGO data. Physics Letters B, 2018, 778: 64-70.
\bibitem{ref6} Allen G, Andreoni I, Bachelet E, et al. Deep Learning for Multi-Messenger Astrophysics: A Gateway for Discovery in the Big Data Era. arXiv:1902.00522, 2019.
\bibitem{ref7} Shen H, Huerta E A, Zhao Z. Deep Learning at Scale for Gravitational Wave Parameter Estimation of Binary Black Hole Mergers. arXiv:1903.01998, 2019.
\bibitem{r2} Chatterjee C, Wen L, Vinsen K, Kovalam M, and Datta A, Using deep learning to localize gravitational wave sources, Phys. Rev. D, 2019, 100, 103025.
\bibitem{addcao} Wang H,  Cao Z J, Liu X L, Wu S C, Zhu J Y, Gravitational wave signal recognition of O1 data by deep learning, Phys. Rev. D, 2020, 101, 104003.
\bibitem{ref26} Wei W, Huerta E A. Gravitational Wave Denoising of Binary Black Hole Mergers with Deep Learning. arXiv:1901.00869, 2019.
\bibitem{ref29} Li X, Yu W, Fan X. A Method Of Detecting Gravitational Wave Based On Time-frequency Analysis And Convolutional Neural Networks. arXiv:1712.00356, 2017.
\bibitem{ref31} Gonzlez J A, Guzmn F S. Characterizing the velocity of a wandering black hole and properties of the surrounding medium using convolutional neural networks. Physical Review D, 2018, 97(6): 063001.
\bibitem{BNN1} Levasseur L P, Hezaveh Y D, Wechsler R H. Uncertainties in Parameters Estimated with Neural Networks: Application to Strong Gravitational Lensing. The Astrophysical Journal Letters, 850:L7 (5pp), (2017)
\bibitem{BNN2} Hezaveh Y D, Levasseur L P, Marshall P J, Fast automated analysis of strong gravitational lenses with convolutional neural networks. Nature 548, p555-557 (2017)
\bibitem{BNN3} Celia E R, et al. A deep learning approach to cosmological dark energy models. arXiv:1910.02788v1
\bibitem{ref30} Burke-Spolaor S, Taylor S R, Charisi M, et al. The astrophysics of nanohertz gravitational waves. The Astronomy and Astrophysics Review, 2019, 27(1): 5.
\bibitem{ref11} Taylor S R, et al. Detecting eccentric supermassive black hole binaries with pulsar timing arrays: resolvable source strategies. The Astrophysical Journal 817.1 (2016): 70.
\bibitem{ref16} Rebei A, Huerta E A, Wang S, et al. Fusing numerical relativity and deep learning to detect higher-order multipole waveforms from eccentric binary black hole mergers. Physical Review D, 2019, 100(4): 044025.
 \bibitem{ref10} Peters P C, and Jon M. Gravitational radiation from point masses in a Keplerian orbit. Physical Review 131.1 (1963): 435.
\bibitem{addpulsar} Shannon R M, et al. Gravitational waves from binary supermassive black holes missing in pulsar observations. arXiv:1509.07320, 2015.
\bibitem{ref12}  George D, Huerta E A. Deep neural networks to enable real-time multimessenger astrophysics. Physical Review D, 2018, 97(4): 044039.
\bibitem{ref13}  George D, Huerta E A. Deep Learning for real-time gravitational wave detection and parameter estimation: Results with Advanced LIGO data. Physics Letters B, 2018, 778: 64-70.
\bibitem{ref14} Fan X L, Li J, Li X, et al. Applying deep neural networks to the detection and space parameter estimation of compact binary coalescence with a network of gravitational wave detectors. SCIENCE CHINA Physics, Mechanics  Astronomy, 2019, 62(6): 969512.
\bibitem{wen56} Miquel L M, Alejandro T F, Jose A F, and Antonio M. Classification of gravitational-wave glitches via dictionary learning. Classical and Quantum Gravity, 2019, 36(7):075005.
\bibitem{wen57} Massimiliano R and Elena C. Image-based deep learning for classification of noise transients in gravitational wave detectors, Classical and Quantum Gravity, 2018, 35, 9.
\bibitem{wen58} Hongyu Shen, Daniel George, E. A. Huerta, and Zhizhen Zhao. Denoising gravitational waves with enhanced deep recurrent denoising auto-encoders, ICASSP 2019 IEEE International Conference on Acoustics, Speech and Signal Processing (ICASSP).
\bibitem{wen59} Marco C, Kai S, and Teerth G. Finding the origin of noise transients in ligo data with machine learning. Commun. Comput. Phys., 2019, 25, 963-987.
\bibitem{wen60} Alejandro T F, Elena C, Antonio M, Jose A F, and Jose M I. Total variation methods for gravitational-wave denoising: Performance tests on advanced ligo data. Phys. Rev. D, 2018, 98:084013.
\bibitem{deeplearning} I. Goodfellow, Y. Bengio, and A. Courville, Deep Learning, MIT Press, 2016.
\bibitem{maxpool} Naranjoalcazar J, Perezcastanos S, Zuccarello P, et al. DCASE 2019: CNN depth analysis with different channel inputs for Acoustic Scene Classification. arXiv:1906.0459, 2019.
\bibitem{fullconnected} Sotoudeh M, Thakur A. Computing Linear Restrictions of Neural Networks. arXiv:1908.06214, 2019.
\bibitem{relu} Schmidthieber J. Nonparametric regression using deep neural networks with ReLU activation function. arXiv:1708.06633, 2017.
\bibitem{batchnorm} Loffe S , Szegedy C . Batch Normalization: Accelerating Deep Network Training by Reducing Internal Covariate Shift. arXiv:1502.03167, 2015.
\bibitem{softmax} Lun X, Jia S, Hou Y, et al. GCNs-Net: A Graph Convolutional Neural Network Approach for Decoding Time-resolved EEG Motor Imagery Signals. arXiv:1907.08487, 2020.
\bibitem{ref15} Kingma D P, Ba J Adam: A method for stochastic optimization. arXiv:1412.6980, 2014.

\bibitem{elman1} Junsong W, Jiukun W, Maohua Z, et al. Prediction of internet traffic based on Elman neural network. Chinese Control and Decision Conference, 2009: 1303-1307.

\bibitem{ref23} Shen H, George D, Huerta E A, et al. Denoising gravitational waves using deep learning with recurrent denoising autoencoders. arXiv:1711.09919, 2017.
\bibitem{ref24} Shen H, George D, Huerta E A, et al. Denoising Gravitational Waves with Enhanced Deep Recurrent Denoising Auto-Encoders, ICASSP 2019-2019 IEEE International Conference on Acoustics, Speech and Signal Processing (ICASSP). IEEE, 2019: 3237-3241.
\bibitem{elman} MATLAB Chinese Forum. MATLAB Neural Network 30 Case Analysis. Beijing University of Aeronautics and Astronautics Press, 2010 ,page170-171.
\bibitem{ref25} Vigeland S J, Islo K, Taylor S R, et al. Noise-marginalized optimal statistic: A robust hybrid frequentist-Bayesian statistic for the stochastic gravitational-wave background in pulsar timing arrays. Physical Review D, 2018, 98(4): 044003.
\bibitem{refbay}  Ellis J A, Cornish N J. Transdimensional Bayesian approach to pulsar timing noise analysis. Physical Review D, 2016,93(8): 084048.






\end{thebibliography}
\end{document}